\documentclass[aps,prl,showpacs,raggedbottom,nobalancelastpage,amssymb,twocolumn,groupedaddress]{revtex4}
\usepackage{graphicx}
\usepackage{amsmath}
\usepackage{amsfonts}
\usepackage{amssymb}
\usepackage{epsfig}
\usepackage{color}
\usepackage{dsfont}

\newcommand{\Tr}[1]{\text{Tr}\left\{#1\right\}}
\newcommand{\ParTr}[2]{\text{Tr}_{#1}\left\{#2\right\}}
\newcommand{\bra}[1]{\langle#1\vert}
\newcommand{\ket}[1]{\vert#1\rangle}

\begin{document}

\title{Information Exchange in Causally Nonseparable Processes}

\author{Gianluca~Francica}
\email{gianluca.francica@gmail.com}
\noaffiliation

\date{\today}

\begin{abstract}
For a system of two parties, the process matrix framework predicts the existence of causally nonseparable structures. We characterize the information exchanged, showing that the total entropy of the two parties acts as a measure for the nonseparability.
\end{abstract}

\maketitle
{\it Introduction --} In our understanding of the world is deeply rooted the idea that events obey a definite causal order, where the causes of events must be sought in past events.
Nonetheless, the process matrix formalism of Oreshkov, Costa and Brukner~\cite{Oreshkov12} has been developed in order to study the most general causal structures compatible with local quantum mechanics for two parties Alice and Bob. The formalism predicts causal structures which are causally nonseparable, i.e. neither Alice comes before Bob nor Bob comes before Alice, nor a mixture thereof. These nonseparable causal structures allow us to find a strategy violating the so-called causal inequalities~\cite{Oreshkov12,Branciard16}.
Anyway, there also exist nonseparable processes which admit a causal model~\cite{Feix16}, i.e. they do not violate causal inequalities (an example is the quantum switch~\cite{Chiribella13}).

In this paper we examine the information exchanged between Alice and Bob in performing a strategy violating the causal inequalities. We find that if the total entropy of Alice and Bob is bigger than any separable process having the same non-signalling part, then the process is nonseparable.

{\it Process matrices --} We consider two systems, Alice and Bob, which are locally described by quantum mechanics. Alice and Bob are given some classical inputs labeled by $x$ and $y$, and return some classical outputs $a$ and $b$, respectively.
By taking in account Alice, for each input $x$ and output $a$, we associate an operation described by a completely positive map $\mathcal M^{A_I A_O}_{a|x}: \mathcal L (\mathcal H^{A_I})\to \mathcal L (\mathcal H^{A_O})$, where $\mathcal L (\mathcal H^{X})$ is the space of linear operators over a Hilbert space $\mathcal H^X$ of dimension $d_X$. We note that all the maps must sum up to a trace-preserving map.
Using the Choi-Jamiol{\l}kowski isomorphism~\cite{Choi75,Jamiolkowski72}, we represent the map $\mathcal M^{A_I A_O}_{a|x}$ as the operator $M^{A_I A_O}_{a|x}=[I^{A_I}\otimes \mathcal M_{a|x}(\ket{\varphi^+}\bra{\varphi^+})]^T$ where $I^{X}$ is the identity matrix on $\mathcal H^{X}$ and $\ket{\varphi^+}=\sum \ket{ii}$. The operators $M^{A_I A_O}_{a|x}$ are such that $M^{A_I A_O}_{a|x}\geq 0$ for each $a$ and $\ParTr{A_O}{\sum_a M^{A_I A_O}_{a|x}}=I^{A_I}$.
Similarly, for Bob we get the operators $M^{B_I B_O}_{b|y}$. The joint conditional probability reads
\begin{equation}\label{cond_prob}
p(a,b|x,y)=\Tr{(M^{A_I A_O}_{a|x}\otimes M^{B_I B_O}_{b|y})W}
\end{equation}
where $W$ is the so-called process matrix, which is an hermitian operator such that the probabilities given by Eq.~\eqref{cond_prob} are non-negative and normalized.
In particular $W$ needs to satisfy the conditions~\cite{Araujo15}
\begin{eqnarray}
W&\geq&0\\
\Tr{W} &=& d_{A_O} d_{B_O}\\
_{B_I B_O} W &=& _{A_O B_I B_O} W\\
_{A_I A_O} W &=& _{A_I A_O B_O} W\\
W &=& _{B_O} W+{_{A_O} W} -{_{A_O B_O}W}
\end{eqnarray}
where we have defined the operation
\begin{equation}
_{X} W = \frac{I^X}{d_X}\otimes\ParTr{X}{W}
\end{equation}
If Bob cannot signal to Alice or Alice cannot signal to Bob we have the process matrices $W^{A\prec B}=W^{A_I A_O B_I}\otimes I^{B_O}$ or $W^{B\prec A}=W^{A_I B_I B_O}\otimes I^{A_O}$, respectively.
A process is causally separable if the process matrix can be expressed in a convex combination
\begin{equation}
W_{sep}=q W^{A\prec B} + (1-q) W^{B\prec A}
\end{equation}
We note that $_{B_0}W^{A\prec B}=W^{A\prec B}$ and $_{A_0}W^{B\prec A}=W^{B\prec A}$. Furthermore, given a process matrix $W$, we can calculate its non-signalling part $\Delta(W)= {_{A_O B_O}W}$. For instance, in the case in which $d_{A_I}=d_{A_O}=d_{B_I}=d_{B_O}=2$, this results by noting that the identity matrix $\sigma_0=I$ and the Pauli matrices $\sigma_x$,$\sigma_y$ and $\sigma_z$ are a basis of the space of the Hermitian operators over $\mathcal H^X$ with $X=A_I,A_O,B_I,B_O$ and the Pauli matrices are traceless. This can be immediately generalized for a dimension $d$, where the basis will be formed by the identity matrix and $d^2-1$ traceless operators.

We recall that causal nonseparability can be inferred by using causal inequalities~\cite{Oreshkov12,Branciard16,Araujo15,Baumeler14,Oreshkov16}. For the case where $x$,$y$,$a$ and $b$ are bits, a causal inequality is a bound on the probability of success of the ``guess your neighbor’s input'' game~\cite{Branciard16}. For uniform input bits $x$ and $y$, the probability of success is $p_{succ}=1/4\sum_{x,y}p(a=y,b=x|x,y)$ and for a separable process $p_{succ}\leq 1/2$, but it is known that there are nonseparable processes such that $p_{succ}>1/2$. Anyway, there also are nonseparable process, having a causal model, that do not violate causal inequalities~\cite{Feix16}.

{\it Information exchange --} We ask how the information exchanged in a nonseparable process behaves. In general, this information will be encoded in the joint probabilities $p(a,b)$, which can be calculated as
\begin{equation}
p(a,b) = \sum_{x,y} p(x,y) p(a,b|x,y)
\end{equation}
where $p(x,y)$ is the probability to have the inputs $x$ and $y$.
In particular, the information in the outputs $a$ and $b$ is quantified by the Shannon entropy $H_{A,B}(W)=-\sum p(a,b) \log_2 p(a,b)$.
By considering the marginal entropies $H_{A}(W)$ and $H_{B}(W)$, defined in a similar way, the correlations between the outputs $a$ and $b$ is measured by the mutual information $I_{A:B}(W)= H_{A}+H_{B}-H_{A,B}$. Conversely, the conditional entropy $H_{A|B}(W)=H_{A,B}-H_{B}$ leads to the equivalent definition of the mutual information $I_{A:B}(W)= H_{A}-H_{A|B}$. We will study how the nonseparability of the process influences these quantities.
We start by noting that $H_{A}(W^{A\prec B})=H_{A}(\Delta(W^{A\prec B}))$ since Bob cannot signal to Alice, and similarly $H_{B}(W^{B\prec A})=H_{B}(\Delta(W^{B\prec A}))$. Thus, we expect that both the marginal entropies for a nonseparable process are different from the corresponding non-signalling case (i.e. Alice and Bob exchange information). Anyway, this also happens for a convex combination of $W^{A\prec B}$ and $W^{B\prec A}$, such that we cannot say if the process is separable or nonseparable.
With the aim to quantify the nonseparability, given a process matrix $W$ we consider the set $\mathcal S_W$ of all the separable processes $W_{sep}$ such that $\Delta(W_{sep})=\Delta(W)$. We find our main result, i.e. there exist nonseparable process matrices $W$ such that
\begin{equation}\label{inequality}
H_{A,B}(W) > \max_{W_{sep}\in \mathcal S_W} H_{A,B}(W_{sep})
\end{equation}
i.e. more information is encoded in Alice and Bob with respect to the separable case. Of course the non-signalling part $\Delta(W)$ belongs to $\mathcal S_W$, then, if the inequality in Eq.~\eqref{inequality} holds, Alice and Bob exchange information via a causally nonseparable structure. We find that a similar condition does not hold for the others quantities $H_{A}$, $H_{A|B}$ and $I_{A:B}$, thus the total entropy $H_{A,B}$ plays a special role in the nonseparability context. Furthermore, we observe that nonseparability does not implies the inequality in Eq.~\eqref{inequality}, since there are nonseparable processes which admit a causal model for which the inequality is not satisfied.

In order to proof our result, we focus on the case $d_{A_I}=d_{A_O}=d_{B_I}=d_{B_O}=2$, and $a$,$b$,$x$ and $y$ are bits. We consider the nonseparable process which violates causal inequalities~\cite{Branciard16}
\begin{equation}
W=\frac{1}{4}\left(I^{\otimes 4} + \frac{\sigma_z^{A_I}\sigma_z^{A_O}\sigma_z^{B_I}I^{B_O}+\sigma_z^{A_I}I^{A_O}\sigma_x^{B_I}\sigma_x^{B_O}}{\sqrt{2}}\right)
\end{equation}
where the tensor products are implicit, and the local operations
\begin{eqnarray}
M^{A_I A_O}_{0|0}&=&M^{B_I B_O}_{0|0}=0\\
M^{A_I A_O}_{1|0}&=&M^{B_I B_O}_{1|0}=\ket{\varphi^+}\bra{\varphi^+}\\
M^{A_I A_O}_{0|1}&=&M^{B_I B_O}_{0|1}=\ket{0}\bra{0}\otimes \ket{0}\bra{0}\\
M^{A_I A_O}_{1|1}&=&M^{B_I B_O}_{1|1}=\ket{1}\bra{1}\otimes \ket{0}\bra{0}
\end{eqnarray}
where $\ket{0}$ and $\ket{1}$ are eigenstates of $\sigma_z$ with eigenvalues $1$ and $-1$, respectively.
Thus, when their input is $0$, Alice and Bob  transmit their incoming state, since $\ket{\varphi^+}\bra{\varphi^+}$ is the Choi-Jamiol{\l}kowski  representation of the identity channel, and output the value $1$. When their input is $1$, Alice and Bob perform a measurement in the $\sigma_z$ basis, whose result defines their classical output, and send out the fixed state $\ket{0}\bra{0}$.
We take $p(x,y)=1/4$, thus the probabilities $p(a,b)$ are $p(0,0)=(1+1/\sqrt{2})/16$, $p(0,1)=p(1,0)=(3+1/\sqrt{2})/16$ and $p(1,1)=(9-3/\sqrt{2})/16$.
In order to calculate $\max H_{A,B}(W_{sep})$ we consider $W_{sep}=q W^{A \prec B}+(1-q)W^{B \prec A}$ where $W^{A \prec B}$ and $W^{B \prec A}$ are parametrized as
\begin{eqnarray}
W^{A \prec B} &=& \frac{I^{\otimes 4}}{4} + \sum c_{\alpha i j} \sigma^{A_I}_\alpha\sigma^{A_O}_i\sigma^{B_I}_j I^{B_O}\\
W^{B \prec A} &=& \frac{I^{\otimes 4}}{4} + \sum c'_{i \alpha j} \sigma^{A_I}_i I^{A_O}\sigma^{B_I}_\alpha\sigma^{B_O}_j
\end{eqnarray}
where $c_{\alpha i j}$ and $c'_{i \alpha j}$ are $72$ real parameters, with $\alpha=0,x,y,z$ and $i,j=x,y,z$. We start by choosing the parameters $q$, $c_{\alpha i j}$ and $c'_{i \alpha j}$ random such that $W^{A \prec B}\geq 0$ and $W^{B \prec A}\geq 0$. Then we calculate the maximum by changing only one parameter at a time, while keeping the other ones fixed. We iterate this procedure until the parameters $q$, $c_{\alpha i j}$ and $c'_{i \alpha j}$ do not change under a certain value. By iterating the algorithm $100$ times, we find the estimation $\max H_{A,B}(W_{sep})\approx 1.795$ which is smaller than $H_{A,B}(W)= 1.845\ldots$.
On the other hand, we consider the process matrix $W$ defined by~\cite{Feix16}
\begin{eqnarray}
W^{A\prec B} &=& \frac{1}{4}\left(I^{\otimes 4}+\frac{1}{3}\sum_i I^{A_I}\sigma^{A_O}_i\sigma^{B_I}_i I^{B_O}\right)\\
W^{B\prec A} &=& \frac{1}{4}\left(I^{\otimes 4}+\sigma^{A_I}_z I^{A_O} \sigma^{B_I}_x \sigma^{B_O}_z\right)\\
W&=&q W^{A\prec B} + (1-q+\epsilon) W^{B\prec A}-\epsilon \frac{I^{\otimes 4}}{4}
\end{eqnarray}
with $\epsilon>0$ such that the process is nonseparable but admits a causal model. We have that $\max_{q,\epsilon}H_{A,B}(W) \approx 1.68$ such that the inequality in Eq.~\eqref{inequality} is not satisfied.

{\it Conclusions --} In summary, we have investigated the information exchanged between two parties with respect to the process matrix framework, which predicts the existence of causally nonseparable structures. We find that, from the point of view of the information exchanged, a separable process is different from a nonseparable one violating causal inequalities. In particular a nonseparable process can be characterized by a total entropy of the two parties which is bigger than the one of any corresponding separable process.

To our knowledge, a characterization of the information of nonseparable processes has not yet been investigated in the literature. In conclusion, beyond their foundational implications, our findings are directly relevant to discern separable and nonseparable processes.

\end{document}